\documentclass[12pt]{article}
\usepackage{graphicx}
\usepackage{amssymb,amsmath,amsfonts,palatino,amsthm}
\usepackage{amssymb}
\usepackage{epstopdf}
\newcommand{\beq}{\begin{equation}}
\newcommand{\eeq}{\end{equation}}

\newcommand{\f}{\begin{equation}}
\newcommand{\ff}{\end{equation}}

\begin{document}

\title{Newtonian gravity in loop quantum gravity}
\author{Lee Smolin\thanks{lsmolin@perimeterinstitute.ca}
\\  
\\
Perimeter Institute for Theoretical Physics,\\
31 Caroline Street North, Waterloo, Ontario N2J 2Y5, Canada}
\date{\today}
\maketitle

\begin{abstract}

We apply a recent argument of Verlinde to loop quantum gravity, to conclude that
Newton's law of gravity emerges in an appropriate limit and setting.  This is possible because the relationship between area and entropy is realized in loop quantum gravity when boundaries are imposed on a quantum spacetime.

\end{abstract}
\newpage

\tableofcontents

\section{Introduction}

The idea that the unification of quantum theory with gravity is essentially thermodynamic has been on the table since the discovery of the laws of black hole thermodynamics\cite{bhlaws} and Bekenstein's discovery of black hole entropy\cite{Bek}.  The discoveries of the Unruh temperature\cite{unruh} and Hawking radiation\cite{hawking} strengthened the reason for hoping for a deep
relationship between gravity, quantum physics and thermodynamics.  

Very early in this history, Bekenstein hypothesized that the entropy of any isolated system is bounded by its area\cite{Bek}.  In 1994 't Hooft extended this to the bold conjecture that the degrees of freedom needed to describe an isolated system in nature can be considered to live on a two surface surrounding the region,with the number of degrees of freedom finite and proportional to the area in Planck units\cite{thooft}.  He called this the {\it holographic principle} and since then we have come to call any application of the relationship between area and entropy as ``holographic."   As developed by Susskind\cite{lenny} and then Maldacena\cite{juan} and many others this led, in the context of string theory and supersymmetric
quantum gauge theory,  to the $AdS/CFT$ correspondence. 

The furthest realization of this idea to date, in the context of gravitational theory,  is the discovery by Jacobson that the Einstein equations can be derived from the laws of thermodynamics, assuming only that Bekenstein's proportionality between area and entropy is universal\cite{Ted}.  This idea has been studied also by Padmanabhan\cite{paddy} and others, but there has remained the feeling that there was a further discovery, just over the horizon.  In a remarkable paper, Erik Verlinde has provided the next step, which is a non-relativistic analogue of Jacobson's argument, in which he derives Newton's law of gravity from thermodynamics plus the relationship between area and entropy\cite{Erik}.  A different argument to the same conclusion has also been provided by Padmanabhan\cite{paddy-NT}\footnote{Verlinde and Padmanabhan's argument both lead to Newton's law but the logic is different.  Both use the equipartition relation, but Verlinde employs the notion of an entropic force, while Padmanabhan gets the gravitational acceleration by inverting Unruh's relation between temperature and acceleration}.

In this paper I show that a version of Verlinde's argument can be run in loop quantum gravity (LQG).  
This strengthens the case for taking LQG as a candidate for a quantum theory of gravity.  Previously it has been shown that the theory has massless  spin two excitations\cite{gravitons}, which have the correct propagators in the low energy limit\cite{propagator}, but there has not before been a direct
demonstration that the theory has a limit which yields Newtonian gravity.  

Indeed, once one sees Verlinde's strategy, its application to LQG is immediate, as it relies on 
a realization of the area-entropy relation within LQG that has been known for many years\cite{linking}.  Before giving the details, it may be helpful to sketch the reasons for this.  

The story goes back to papers of Crane\cite{louis-holo}, which anticipated 't Hooft's formulation of the holographic principle.  Crane proposed that in quantum cosmology,
Hilbert spaces should be associated not with the whole universe, but with any choice of
a boundary that splits the universe into two parts.  The idea was that the
observers and their measuring instruments live on one side of the boundary, and they observe
the quantum gravity dynamics on the other side by means of measurements made on the boundary, and recorded in a boundary Hilbert space.  

Crane proposed that these boundary Hilbert spaces be constructed from topological quantum field theory (to be specific, Chern-Simons theory, as the boundaries are $2+1$ dimensional.)  This was for two reasons.  First, the Hilbert spaces associated with topological quantum field theories are finite dimensional, so that the relationship between area and entropy can be naturally implemented.  Second, if one considers different  boundaries related by cobordisms within the space, the Hilbert spaces on them should be mapped to each other by categorical relations that are naturally satisfied in topological quantum field theories.  

This was an intriguing suggestion because  there is a close connection between topological field theories and the dynamics of general relativity, which is summarized in the next section.   
Topological field theories play a role for diffeomorphism invariant quantum theories similar to that  played by harmonic oscillators in conventional quantum field theories.  They provide simple solvable examples and, more than that, they provide the Hilbert spaces on which the full non-linear dynamics acts. This deep relation is the basis for the Plebanski action principle, and underlies the Ashtekar formulation of canonical dynamics\cite{abhay} and the spin foam formulations of path integral quantization\cite{barrett-crane,KL}.

Inspired by Crane's proposal,  it was shown in \cite{linking} how a holographic formulation of quantum gravity necessarily emerges within LQG when certain boundary conditions are imposed.  The role of Chern-Simons theory in the construction of boundary Hilbert spaces and its observables was worked out in detail and it was shown that Bekenstein's area-entropy relationship was a necessary consequence. 
It was then noticed\cite{kirill} that similar boundary Hilbert spaces constructed from Chern-Simons theories are required to describe horizons of black holes\cite{isolated}.  This led to enormous 
progress in the understanding of black hole entropy and the quantum dynamics of 
horizons\cite{abhay-isolated}.  

To summarize, the holographic character of quantum gravity follows from the close relationship of the dynamics of general relativity to topological field theories, both classically and quantum mechanically.  The relationship to TQFT provides the Hilbert spaces for boundaries and horizons, which being finite dimensional turn out to realize directly the relationship between area and entropy. 

The fact that LQG is deeply tied up with the holographic principle was thus known when Jacobson published \cite{Ted}, and it has since then been a goal to apply his argument to LQG.  Now that Verlinde has shown the way, it is straightforward to do this, at least in the non-relativistic approximation.  

In the next section we review the realization of holography in LQG.   I summarize only what is needed for the application to Verlinde's argument, which is given in section 3.  More details are in \cite{linking}.  The paper then closes with some comments.

\section{The holographic character of loop quantum gravity}


To bring out the holographic character of loop quantum gravity,  we will think about quantum gravity experiments in the following way.   
We consider a spacetime 
$\cal M $ of topology $\Sigma \times R$ for some three topology $\Sigma$.  We assume
that $\Sigma$  can be divided into two parts 
$\Sigma= \Sigma_{interior} \cup \Sigma_{exterior}$, with a boundary between them given by 
a spatial two surface $\cal S$.  This division can be extended to spacetime so $\cal M$ decomposes into the union of 
two regions ${\cal M}_{interior}$ and ${\cal M}_{exterior}$ joined at a boundary  ${\cal S} \times R$.

We will assume that the internal region is to be described by a quantum gravity theory, while
the external region describes a world with classical spacetime and quantum matter, which is
where our observers live.  We study quantum dynamics only in the interior region, so for 
simplicity we assume that the external region is a piece
of deSitter or antideSitter spacetime with a cosmological constant, $\Lambda$.  We assume also that the boundary is the site of measurements that observers in the exterior region make of the
quantum gravity state and dynamics in the interior region.  These measurements are  recorded in a suitable Hilbert space associated
with the boundary.  

{\it Remarkably, the assumption that the external region is a piece of deSitter or anti-deSitter space-time provides exactly such a boundary Hilbert space.}

This comes about in the following way. (A)deSitter spacetime is characterized by the following
condition
\f
F^{+ab} = \frac{\Lambda}{3} \Sigma^{ab} 
\label{dS}
\ff
where $F^{+ ab}$ is the self-dual part of the curvature two-form and 
$\Sigma^{ ab} $ is the self-dual two form constructed from the frame fields
$e^a$ as the self-dual part of $e^a \wedge e^b $.  ($a,b=0,...,3$ are internal lorentz indices. )

This implies that on the spatial boundary $S$ the following condition is satisfied for two forms
pulled back to the two dimensional boundary:
\f
 \epsilon_{ijk}  e^j \wedge e^k = \frac{k}{2\pi} G F^{+}_i   
\label{dS2}
\ff
where $e^i$ (with $i=1,2,3$) are the spatial frame field one-forms pulled back into the boundary.  (We use the fact that the self-dual part of the Lorentz algebra is valued in $SU(2)$.) Here $G$ is Newton's constant and $k$ is
\f
k=\frac{6\pi}{\hbar G\Lambda} 
\label{kdef}
\ff

When we describe the quantum spacetime in the interior region we have to impose 
(\ref{dS2}) as a boundary condition on $\cal S$ in order to  match the components of forms pulled back onto the boundary from  each side.  

 {\it This results necessarily implies the emergence of a Hilbert space on the 
boundary $\cal S$, which automatically satisfies the Bekenstein relation between area and
entropy} (defined as the log of the dimension of the boundary Hilbert space.) 

This comes about as follows.  As first discovered by Plebanski\cite{Pleb}, an action principle for general 
relativity exists which is of the form
\f
S^{Pleb} =  \frac{1}{G} \int_{\cal M}  B^i \wedge F_i +....
\label{Pleb1}
\ff
Here the three $B^i$ are two forms valued in the lie algebra of $SU(2)$ and 
the terms left out are {\it non-derivative terms in $B^i$ and lagrange multiplier fields.} 
$F^i$ is the field strength of an $SU(2)$ connection $A^i$ which in the end turns out to 
be the left-handed part of the space-time connection.

The relationship to topological quantum field theory arises because if we restrict the 
Plebanski action (\ref{Pleb1}) to the term containing derivatives, we find 
\f
S^{BF} =  \frac{1}{G} \int_{\cal M}  B^i \wedge F_i 
\label{BF}
\ff
which is a topological field theory.  Since the commutation relations and path integral measure reflect information only in the derivative terms of an action, the quantization of general relativity is
closely related to that of this topological field theory.  Since the latter can be quantized completely and rigorously, this gives us the basis for quantizing general relativity or any of a large number of other gravity theories that have an action of the form of (\ref{Pleb1}).  

For the description of the situation we have in mind, the quantum dynamics is applied in the
interior region only, so we need to consider the action (\ref{Pleb1}) in the interior region.
In this case a boundary term must be added to the action to cancel 
a boundary term produced by the variation of the action by the connection.  
The boundary condition we want to impose is (\ref{dS2}) (possibly extended also to the space-time
boundary\footnote{This depends on whether we are dealing with the Euclidean or Lorentzian
case, see \cite{hologr} for a detailed discussion.}).  The only boundary term that can be
added to the action compatible with the imposition of (\ref{dS2}) as a boundary condition
is the Chern-Simons invariant of the connection $A^i$, $Y(A)_{CS}$.
\f
S^{Pleb} =\frac{1}{G} \int_{\cal M} B^i \wedge F_i +....  -\frac{k}{4\pi}  \int_{{\cal S} \times R} Y(A)_{CS}
\label{Pleb2}
\ff
It is easy to verify the compatibility and consistency of the equations of motion gotten by
variation of this action with the boundary conditions (\ref{dS2}) that indicate that the boundary
joins onto (A)dS spacetime.    The key point is that on solutions $B^i$ are equal to the self-dual two forms $\Sigma^{i}. $

The task of carrying the quantization of the theory in the presence of these boundary conditions and terms was studied in \cite{linking} and {\it a  key result was that  Bekenstein's area-entropy relation is necessarily satisfied.}  This was obtained as follows.  

On $\cal S$ there is an area operator\cite{area} $\hat{\cal A}_{\cal S}$, 
whose spectrum is completely understood\cite{area-rigorous}.  $\hat{\cal A}_{\cal S}$ has
a discrete spectrum given by choices of $N$ punctures on $\cal S$ and representation labels
$j_i$, $i=1,...N$.   The $j_i$ are spins, that is representations\footnote{More precisely, its quantum deformation at level $k$, but we will ignore this detail as we want in the end to 
take $k \rightarrow \infty$.} of $SU(2)$.  The Gauss's law constraint turns out to require  that each puncture, $p_i$,  joins to an edge $e_i$ of a spin network in the interior,  labeled by representation $j_i$\footnote{There is a more general case in which
$n_i$ may be a vertex with valence $v_i >2$ with edges running in $\cal S$ but we will assume
that the surface has been chosen to avoid this, as the presence of a node of  a graph in a surface
is of measure zero.  Nonetheless everything below extends immediately to this 
case\cite{area-rigorous}.}.

The area of $\cal S$, which is the eigenvalue of $\hat{\cal A}_{\cal S}$, depends on the $j_i$'s and is
given for these cases by\cite{area}
\f
a (j_i) = 8 \pi \frac{\hbar G}{c^3} \gamma \sum_i \sqrt{j_i (j_i+1)}
\label{area1}
\ff
Here $\gamma$ is a parameter analogous to the
$\Theta$ parameter of $QCD$ called the {\it Immirzi} parameter.  

Part of the specification of the experiment is to fix the induced
metric $r_{\alpha \beta}$ on the boundary, which fixes also the
total area
\f
A=\int_{\cal S} \sqrt{r}.
\ff  
We assume then that the punctures are picked so that
$a(j_i) \approx A$.  We also assume that they are randomly distributed on $\cal S$ according to the measure
given by $\sqrt{r}$.

When this boundary conditions is satisfied, there is induced, on the boundary, a set of  Hilbert
spaces.  These  are used to record information gotten by measurements
that observers in the external region make on the boundary of the degrees of freedom of
the quantum gravitational field in the interior.  

Using the spectrum of $\hat{\cal A}_{\cal S}$ the diffeomorphism invariant 
Hilbert space ${\cal H}^{diffeo}$
can in this case be decomposed as follows.
\f
{\cal H}^{diffeo}= \bigoplus_{N=1}^\infty \bigoplus_{j_i, ... , j_N} {\cal H}_{j_i, ... , j_N} ^{bulk,diffeo} \otimes
{\cal H}_{j_i, ... , j_N} ^{boundary} 
\ff
Here ${\cal H}_{j_i, ... , j_N} ^{boundary} $ is the finite dimensional Hilbert space of $SU(2)$
Chern-Simons theory at a level $k$ on the punctured two surface $\cal S$.  

The bulk Hilbert space ${\cal H}_{j_i, ... , j_N} ^{bulk,diffeo} $ has a countable basis given by
diffeomorphism classes\footnote{with diffeomorphisms fixed to the identity on the boundary.} of spin networks  $\Gamma$ in the interior that join to the boundary, such that 
$\Gamma$ has $N$ edges $e_i$ with spins $j_i$ that meet the boundary at the 
punctures, $p_i$.  All the states of ${\cal H}_{j_i, ... , j_N} ^{bulk,diffeo} $ are in the kernel
of the generators of spatial diffeomorphisms ${\cal D}(v)$ for vector fields in $\Sigma$ that
vanish on $\cal S$.  They are also in the kernel of a Gauss's law operator, $\hat{\cal G}$ that includes
contributions from both bulk and boundary terms which is not restricted to vanish on
$\cal S$\footnote{For details of the definitions of these operators and their kernels 
see \cite{linking}.}  The Gauss's law constraint hence links the information in the bulk state with the information in the boundary state.  

The Hamiltonian constraint operator $\hat{\cal C} (N)$ acts on 
${\cal H}_{j_i, ... , j_N} ^{bulk,diffeo} $ with
lapse $N$ vanishing on $\cal S$.  It's kernel is the space of physical bulk states
 ${\cal H}_{j_i, ... , j_N} ^{bulk,phys} $ so the space of physical states is given by
 \f
{\cal H}^{phys}= \bigoplus_{n=1}^\infty \bigoplus_{j_i, ... , j_N} {\cal H}_{j_i, ... , j_N} ^{bulk,phys} \otimes
{\cal H}_{j_i, ... , j_N} ^{boundary} 
\ff
There is a nonvanishing Hamiltonian operator $\hat{H}$ derived from the classical hamiltonian 
$H_{\cal S}$.  Given a choice of a lapse $N$ that {\it does not vanish} on $\cal S$ the
classical Hamiltonian is given by
\f
H(N) = \int_{\cal S} h N + \int_\Sigma  \left (
N {\cal C} + {\cal D}_a v^a + {\cal G}_i \rho^i \right  )
\ff
where $v^i$ is an arbitrary vector field vanishing on $\cal S$ and $\rho^i$ is a parameter in the
generators of $SU(2)$.  When the constraints are satisfied the Hamiltonian then reduces to a boundary term given by $h$.  
 In the ordinary variables
these correspond to the Brown-York hamiltonian\cite{brown-york}
\f
h =\sqrt{q}K 
\ff
where $K$ is the extrinsic curvature. In a limit in which the surface
$S$ is taken to infinity in the asymptotically flat case it goes over into the $ADM$ mass. 
In the Ashtekar formulation  (with the lapse a density of weight minus one)
\f
\int_{\cal S} N h = \int_{\cal S} d^2\sigma_a N {\cal R}e \left ( A_{bi} \tilde{E}^a_j \tilde{E}^b_k \epsilon_{ijk} \right ) 
\ff
Different forms for $h$ have been proposed and their quantizations studied in
loop quantum gravity\cite{linking,thomas-boundary,yongge-boundary}.

What we need for the following is only that in all these 
cases\cite{linking,thomas-boundary,yongge-boundary} the quantum Hamiltonian acting
on the physical Hilbert space takes the form
\f
\hat{H} = \sum_i \hat{h}_i
\ff
where $\hat{h}_i$ is an operator that acts at the puncture $p_i$.  

The entropy relevant for observations made on the boundary $\cal S$ can be defined by the
log of the dimension of ${\cal H}_{j_i, ... , j_N} ^{boundary} $.  
\f
S_{j_i} \equiv  \ln \mbox{dim} {\cal H}_{j_i, ... , j_N} ^{boundary} 
\label{Sdef1}
\ff

The dimension of ${\cal H}_{j_i, ... , j_N} ^{boundary} $  is given approximately for large $k$ by
$\prod_i (2j_i + 1 ) $ .  Given an area $A$ large in Planck units, the largest dimension of 
${\cal H}_{j_i, ... , j_n} ^{boundary} $, such that $a (j_i)$ is close to $A$, is gotten by choosing\footnote{Some authors here take an alternative strategy of finding the expectation value of
area by averaging over all representations, this gives a slightly different value of the Immirzi parameter, but does not affect the following.} all the $j_i = \frac{1}{2}$.   
When entropy is maximized given $A$ we can then approximate the entropy (recall that $k$ is assumed large)
\f
S_{A} \equiv  \ln \mbox{dim} {\cal H}_{\frac{1}{2}^N} ^{boundary} \approx  N \ln{2}
\ff

When this is the case we have, using (\ref{area1}), the relationship between area and entropy 
\f
S_{A} = \frac{A c^3}{  4 G \hbar  }
\ff
To get the correct coefficient determines a choice for the Immirzi parameter, $\gamma = \frac{\ln (2) }{ \pi \sqrt{3}}$.

The hypothesis that gravity is fundamentally thermodynamic is essentially that we take 
this entropy seriously as thermodynamic entropy, that is as a measure of disorder. 
In the context of this discussion the microscopic degrees of freedom relevant for
observations of quantum gravitational dynamics are those on the boundary.  These boundary degrees of freedom are hypothesized to be maximally disordered so that their thermodynamic entropy is equal to the log of the dimension of their
Hilbert spaces, given by (\ref{Sdef1}).  

If we want to understand this more deeply, we can think that the classical geometric description is a kind of coarse graining and that the information lost by doing so is measured by the area of a surface which bounds the region so described.  This might be proportional to area because it is a kind of entanglement entropy measuring correlations between the states of quantum geometry on both sides lost when one side is coarse grained to a classical description.

\section{Derivation of Newton's law of gravity}

We now have the ingredients necessary to run a version of Verlinde's argument.  
We consider a situation where $\cal S$ can be considered to be a two-sphere of
a given radius $R$, and we posit that 
there is a spherically symmetric mass distribution in the
interior region, which is approximately static and in equilibrium.  We will study a 
process whereby some small excitation initially in the interior region moves out to the
exterior region, where it is interpreted as a massive particle.  Using the laws
of thermodynamics together with the relations discussed above we show that there is
necessarily an attractive force on that particle which is given by Newton's law of gravitation.  
A necessary step in the derivation, as we will see, is to take the limit $\hbar$ and
$\Lambda$ to zero while taking $c\rightarrow \infty$.  

We then assume that the bulk state corresponds to a spherically symmetric mixed state, so that
the total mass is given by
\f
< \hat{H} > = M c^2
\ff
and the induced metric on $\cal S$, $r_{\alpha \beta}$ is that of a two sphere with area 
$A= 4\pi R^2$ which is large compared to the Planck area and the Schwarzchild radius of $M$.
Since the state can be assumed to maximize entropy as measured by observers at the
boundary we can assume that all the punctures have spin $\frac{1}{2}$ so that there
are $N$ punctures given by
\f
N=\frac{A c^3 }{4 \pi \gamma\hbar G  \sqrt{3}} = \frac{A c^3 }{4 \hbar G \ln (2)} 
\label{Ndef}
\ff

Since the state is spherically symmetric we can assume a version of the equipartition
of energy, so that each node contributes the same amount of energy, 
\f
< \hat{h}_i > = \frac{M c^2}{N}
\ff
Now we are going to consider a process in which an excitation\footnote{Excitations of spin newtworks which can be interpreted as particles have been studied in \cite{braids}.} moves from the interior
of $\cal S$ to the exterior.  We will follow Bekenstein\cite{Bek} in asking how this can be done
to cause the minimal possible perturbation to the system as a whole.  

The process of moving an excitation to the exterior requires the addition of at least one puncture to the boundary.    Microscopically this is required to give meaning to the excitation having 
been moved from the interior to the exterior.   There is then a change
in the area of the surface whose minimal value is
\f
\Delta A = \frac{  \hbar G}{c^3}4  \ln (2)
\ff
There corresponds a minimal change in entropy 
\f
\Delta S =  \ln (2)
\ff
We use capital $\Delta$'s to emphasize that these changes are small, but finite, due to the 
discreteness of the area spectrum and the fact that the Hamiltonian acts at punctures.  

The essence of Jacobson's idea\cite{Ted} is that any such translation of an excitation across the boundary involves a change both of energy and the entropy.  The latter implies a change of the area of the boundary.  In Jacobson's derivation this turns out to imply the Einstein equations. 

A key element introduced by Verlinde \cite{Erik} is that there also must be a temperature associated to the process\footnote{Jacobson got the value of the temperature from the Unruh formula.  Verlinde very cleverly assumes less, gets the temperature from only an 
assumption about the equipartition of energy, and {\it derives} the Unruh formula relating
acceleration to temperature.  In \cite{paddy-NT}, Padmanabhan also defined temperature in terms of equipartition of energy.}, because any  $\Delta U$ is accompanied by a $\Delta S$. 
We must define the temperature $T$ by the usual formula,
\f
\frac{1}{T}= \frac{\partial S}{\partial U}
\ff
Since the translation of the excitation is associated with a change in entropy this means that the
temperature must be finite.  This is a direct consequence of taking the entropy of the boundary seriously as a measure of disorder.  We will determine its value shortly by the equipartition of energy. 
 

If it makes sense to define a temperature then the first law of thermodynamics must be satisfied in this process.  As a result, in addition to the transfer of its own energy, the translation of the excitation from the interior to the exterior must be accompanied by a change in energy
corresponding to the change in entropy.  
\f
\Delta U = T \Delta S
\label{first law}
\ff

Now, once it is external to $\cal S$, the excitation is moving in a classical spacetime. We assume that then it can be
described as a particle.  Because of the quantum
mechanical uncertainty principle there is a minimal distance $\Delta x$ that this particle must be
from $\cal S$ to be reliably considered to be in the external region.  Then the change
in energy $\Delta U$ corresponding to this motion over a distance $\Delta x$ implies there is a
force $F=\frac{\Delta U}{\Delta x}$ acting on the excitation.  By the first law of thermodynamics, this force is given by
\f
 F \Delta x = \Delta U = T \Delta S
\label{firstlaw}
\ff

The definition we have used of $T$ implies the equipartition of energy, which we have already assumed.  
So we write $ M c^2 = \frac{1}{2} {N} T $ which gives us,
\f
T = \frac{2M c^2   }{  N}
\label{Tdef}
\ff
If we now use (\ref{firstlaw}) and (\ref{Ndef}) we have
\f
F= \frac{G M}{R^2} \left (  \frac{\hbar}{\Delta x c}  \right ) 
\left (  \frac{2( \ln(2))^2}{ \pi}    \right ) 
\ff

Now we are interested in taking the non-relatistic classical limit to describe the force in the
external region, so we consider taking $c\rightarrow \infty$ and $\hbar \rightarrow 0$.  
There can only be a non-zero force if we can define a parameter $m$, with dimensions of mass,  that characterizes the excitation so that, with $f$ a dimensionless parameter of order unity, 
\f
m  = \frac{\hbar f }{\Delta x c}
\ff
This tells us that $\Delta x$ must be approximately  the Compton wavelength of a particle with mass $m$.  The proportionality is given by a fudge factor $f$, which we adjust to make $m$ exactly
into the passive gravitational mass\footnote{Another way to say this is that we use $f$ to adjust units for passive gravitational mass so that they are the same as inertial mass.}. 


We choose to set 
\f
f= \frac{2(\ln(2))^2}{  \pi}
\ff 
so that we arrive finally at Newton's law of gravity, 
\f
F= \frac{G M m }{R^2} .  
\ff

We note that the need for a fudge factor is not surprising because the Compton wavelength gives us only a rough value for how high we must translate the particle above the surface $\cal S$ to be considered securely in the exterior.  The important thing is that $f$  is independent of the excitation, so we get the
universality of the gravitational force, as well as its proportionality to the mass\footnote{Another place a fudge factor might appear is in the count of the number of degrees of freedom used in the
definition of the temperature through equipartion; here we have assumed that each puncture can be
counted as one degree of freedom, but the precise number of independent degrees of freedom  per node might depend on the exact form of the quantum
hamiltonian.}.  

We note also that we have not needed the cosmological constant, apart from its role in setting up the 
calculations, so we can take $\Lambda \rightarrow 0$ at the end .   
Finally, for consistency we can check that in this process $\frac{\Delta T}{T} \rightarrow 0$ as 
$\hbar \rightarrow 0$ and $c \rightarrow \infty$ so that it is consistent to treat the process at constant temperature.

\section{Comments and conclusions}

\begin{itemize}

\item{} It is important to emphasize that I have not shown here that classical spacetime emerges from loop quantum gravity, as we have assumed
that there is a classical spacetime in the exterior region where we make measurements.  What has been shown is that {\it if} there is a classical spacetime that emerges {\it then} Newton's law of gravity is  
necessarily  satisfied.

\item{}{ The temperature $T$ is proportional to the  Unruh temperature.}  From (\ref{Tdef}) and (\ref{Ndef}) we see that if
$a=F/m$ is the acceleration of the particle in the external region we have
\f
T = a \frac{\hbar}{\Delta S  c} = T_{Unruh}\frac{2 \pi}{\Delta S}
\ff
Thus,  the Hilbert space of the boundary is hot, when
there is a gravitational field arising from an energy in the bulk.  Indeed, this can be seen
directly from the first law (\ref{firstlaw}).  

\item{}As Verlinde emphasizes, the gravitational force is an entropic force.  It arises because it takes energy to effect an increase in entropy in a system at a finite temperature.  We want to then ask how the equivalence principle could still be true? Here, perhaps, is a simple answer: in a reference frame that is stationary with respect to the excitation, it will remains on the same side of the boundary that defines the reference frame of that observer, so there can be no change in entropy and hence no force.  

\item{}It should be emphasized that the sign comes out right.  The increase
in energy $\Delta U$ on the particle moving to the exterior of $\cal S$ is positive because the change in the entropy is positive.  This means that there is an attractive force pulling the
excitation back towards the mass $M$ in the interior.  So gravity is attractive!

\item{}The parameter $m$ which emerges to characterize the excitation in the external region
is the {\it passive gravitational mass}.  We have to do more to derive relationships to inertial
mass and active gravitational mass.  If we think of the Compton wavelength as a measure of the inertial mass, what we have derived here is the proportionality of that to passive gravitational mass, but not the exact identity. 

\item{}Given that we have found that a finite temperature characterizes the quantum geometry with non-zero mass, it would be desirable to continue to the Euclidean theory to implement the finite
temperature partition function as a path integral in Euclidean quantum gravity.  This can be done directly in the formulation studied here. One finds that in the Euclidean case $k$ becomes an integer because the connection $A^i$ is now real so the group is compact.  One further finds that the temperature of deSitter spacetime can naturally be recovered\cite{wchopin,positive}.  

\item{}We can stress the generality of these results, which follows from the weakness of the assumptions.  The Plebanski type of formulation of general relativity, which is the basis of everything contained here, is extremely general.  It has been extended to supergravity in four and eleven dimensions\cite{super}, to general higher dimensions\cite{higher}, and is the basis for a unification of general relativity with Yang-Mills and Higgs theories\cite{unification}.  There is a large class of extended theories of gravity which can be constructed
by extending the Plebanski action\cite{extended}.  Futhermore, what we use from loop
quantum gravity is very weak, no detailed assumptions about the dynamics are used.  


\item{} It is intriguing to wonder whether the transition between quantum geometry and classical geometry might be dynamical and to have taken place in the very early universe.  This has been
hypothesized in \cite{fotinietal} under the name of {\it geometrogenesis}.  It can be shown that the assumption that the transition is ``holographic", so that the specific heat of a region is proportional to its area rather than its volume, leads directly to scale invariant
fluctuations of the kind seen in the $CMB$\cite{JCL}.

\item{}It is intriguing that, as in the $AdS/CFT$ correspondence, we seem to need a non-zero cosmological constant to set up a holographic description of gravity, even if we here can take the limit $\Lambda \rightarrow 0$ at the end.  

\item{}Note that what is needed to derive Newton's law of gravitation is only a weak  form of the holographic principle\cite{weakholo}. We need only that when we divide a quantum spacetime into two, there is a quasi local description of the interior from the boundary using a boundary Hilbert space in which {\it some} of the observables are defined.   When the quantum constraints are satisfied, one of these will be the Hamiltonian.  The fact that the Hamiltonian is a function only of observeables on the boundary is a consequence only of diffeomorphism invariance.  The other ingredient needed is the relationship between entropy and the area of the boundary, and this follows also directly from basic principles of the theory.  We do not need to assume that all the degrees of freedom of the theory ``live" in the boundary hilbert space, nor that there is a dual theory which is defined only in the boundary.  

\item{}Nonetheless, it is intriguing to wonder if the relationship between area and entropy is even more fundamental than the notion of geometry itself.  Could there be a more fundamental picture, before spacetime emerges in which area has the fundamental meaning of the capacity of a quantum channel by which information flows\cite{quantumholo}?

\item{}These considerations bring to mind an old proposal that the distinction between quantum and thermal statistics is fundamentally related to the principle of inertia\cite{me-qtf}.  Presently in physics we define the vacuum of Minkowski spacetime so that the preferred frames in which the vacuum is at zero  temperature are the same as the preferred frames in which particles with no forces acting on them move in straight lines.  The question is whether this is more than a coincidence and is to be explained by a deep connection between thermodynamics, quantum physics and the principle of inertia.  

\end{itemize}

There is a final remark I would like to close on.  
Verlinde emphasizes that his derivation supports the view that gravitation is an emergent phenomena, so there need be no fundamental degrees of freedom associated with the geometry of spacetime.  There is indeed much reason to be hopeful about this viewpoint and its recent
developments\cite{emergent}.

At the same time, I would like to emphaze that even if the fundamental degrees of freedom do come  from quantizing general relativity or another diffeomorphism invariant
theory, emergent gravity is still necessary. The quantum geometry degrees of freedom, as described in various background independent approaches, such as loop quantum gravity, spin foam models, dynamical causal triangulations, causal sets, etc, do not live in a classical spacetime, that is what is supposed to be good about them.  But we do nonetheless appear to a good approximation to live in a classical spacetime and the burden on any of these approaches is to explain why we do. 

This is well appreciated, and there is much recent progress in the context of spin foam models as well as causal dynamical triangulations towards it.  These come from detailed studies of propagators and measures in path integrals\cite{propagator}. 
However most of these approaches are "brute force" in that they 
depend on the details of the microscopic degrees of freedom and quantum dynamics. 
Even when they work they do not yield clear insight into why they work. The wonderful thing about the arguments of Jacobson and Verlinde is they give a deep reason for why a quantum theory of gravity should yield the phenomena of gravitation.  The reason is thermodynamic and directly a consequence of the holographic relationship between entropy and area.

\section*{ACKNOWLEDGEMENTS}

I would like to thank Laurent Freidel, Sarah Kavassalis, Carlo Rovelli and Simone Speziale  for very helpful comments on a draft of the manuscript.  I am also very grateful to T.Padmanabhan  and  Erik Verlinde for correspondence and Xiao-Gang Wen for conversation on this topic.  
Research at Perimeter Institute for Theoretical Physics is supported in part by the Government of Canada through NSERC and by the Province of
Ontario through MRI.

\end{document}